\documentclass[twocolumn, resetfootnote, trackchanges]{aastex701} %, linenumbers
\definecolor{trackchange}{cmyk}{0.5,0,0,1} %black-blue

\usepackage{graphics,epsf}
\usepackage[utf8]{inputenc}
\usepackage{amsmath}                % American Mathematical Society package
\usepackage{amsfonts}               % American Mathematical Society fonts
\usepackage{amssymb}                % American Mathematical Society symbol
\usepackage{epsfig}                 % EPS figures
\usepackage{graphicx}               % Required for inserting images
\usepackage{float}
\usepackage{color}
\usepackage{multirow}               % double row table entries
\usepackage{hyperref}

\hypersetup{
    colorlinks=true,
    linkcolor=red,   
    urlcolor=cyan}

\usepackage{caption} % Side-effect to manually importing caption: triggering "LaTeX Error: Command \longtable* already defined."
\usepackage[colorinlistoftodos]{todonotes}

\newcommand{\s}{{~\rm s}}

\newcommand{\g}{{~\rm g}}
\newcommand{\G}{{~\rm G}}
\newcommand{\K}{{~\rm K}}

\begin{document}

\title{Quantifying Symmetry: Transformation Information for Planetary Nebulae and Supernova Remnants}
\date{December 2025}

\author[0000-0002-9444-9460]{Dmitry Shishkin}
\affiliation{Department of Physics, Technion - Israel Institute of Technology, Haifa, 3200003, Israel}
\email{s.dmitry@campus.technion.ac.il}
\correspondingauthor{Dmitry Shishkin}

\author[0000-0002-1361-9115]{Amir Michaelis}
\affiliation{Department of Physics, Technion - Israel Institute of Technology, Haifa, 3200003, Israel}
\email{amir.m@aminor-tech.com}

\begin{abstract}
We present a quantitative symmetry-identification pipeline for astrophysical images based on Transformation Information (TI), an information measure of self-similarity under geometric transformations. TI is expressed as a Kullback-Leibler (cross-entropy) divergence between an image and its rotated or reflected counterpart on the overlapping domain. By scanning rotation angles and reflection axes, we obtain TI curves whose local minima identify symmetry operations. We validate the method on a wind-rose pattern and then apply it to planetary nebulae, where the recovered axes trace bipolar and multipolar lobes consistent with morphology-based classifications. Applying TI to supernova remnants yields estimate axes associated with protrusions, rims, and substructure.
To emphasize global morphology, we introduce a thresholded two-level variant that compares binary silhouettes and can reveal outline-driven symmetries. Finally, we quantify symmetry using a minima prominence-to-width score and show that this compact descriptor separates Type Ia and core-collapse remnants into distinct populations for an X-ray sample. TI provides a non-parametric, reproducible framework for symmetry identification, classification and population studies.
\end{abstract}

% https://astrothesaurus.org/concept-select/
% \keywords{supernovae: general -- ISM: supernova remnants}

% ================================================
\section{Introduction}
\label{sec:Introduction}
% ================================================
Many physical, biological, and engineered systems can be understood through the lens of symmetry (e.g., \citealt{BrandingCastellani2003_symmetry, Livio2012_symmetry}). The quantitative and systematic identification of symmetries in physical systems is a central organizing principle, used to classify and deduce the mechanisms that generated a structure and the forces that continue to shape it (e.g., \citealt{Zabrodsky_etal1995_Symmetry,Ierushalmi_etal2020_BioSym}).

\subsection{Symmetry in Astrophysics}
In astrophysics, symmetry plays a central role. Morphology can reveal the dynamics, common shaping mechanisms and evolution of systems across all scales.
In galaxies, structural symmetry can be used as a quantitative morphological metric (e.g. \citealt{Conselice1997_GalaxySym}). Shaping by jets emitted from active galactic nuclei (AGNs), and by extension galaxy clusters, hold details for their shaping mechanisms, activity timeline and interaction dynamics (e.g., \citealt{Horton_etal_2025_AGNmorph}).
In planetary nebulae (PNe), deviations from spherical symmetry in the form of multipolar or point-symmetric structures originate from binary interaction, bi-polar outflow (jet-launching) or variable mass-loss episodes that shape the aforementioned structures (e.g., \citealt{Soker1990_AnsaeFormationPNe, BalickFrank_2002_PNeShape}).
Identification of similarity in astrophysical systems using neural network tools (e.g.,  \citealt{Stoehr_etal2025_morphologyAI}) emphasizes the necessity for quantitative symmetry identifying tools.

Quantifying symmetrical features, or lack thereof, in supernova remnants can have implications for understanding the explosion mechanism and the underlying physical processes shaping the ejecta, both for core-collapse supernovae remnants (CCSNRs, e.g. \citealt{Gabler_etal_2021_CCSNR_structure, Orlando_etal_2025_1987AejectaSim, Braduo_etal_2025_JJEM_structure}) and thermonuclear remnants (Type Ia, e.g., \citealt{Soker2025_SNIP0509, Michaelis_etal2025_Type1aSim, Das_etal2025_TypeIaExp}). 

This prevalence of symmetry-breaking suggests that symmetry-shaping processes are fundamental to understanding astrophysical systems' evolution.
Morphological similarities between apparently unrelated astrophysical systems can deepen our understanding whenever they are potentially governed by a common shaping process: PNe and galaxy clusters (e.g., \citealt{SokerBisker2006}), SNRs and PNe (e.g., \citealt{Soke2024_PNeMorphCCSN}) and SNRs and galaxy clusters cooling flows (e.g., \citealt{Soker2024_SNeMorphCoolingFlows}) or specific galaxies (e.g., \citealt{SokerShishkin2025_W49BMorphCygA}).

Quantifying symmetry (or lack thereof) provides a data-driven avenue to probe system geometry.

\subsection{KL divergence\footnote{A note on terminology. We refer to our distance parameter, applied on image intensity, as KL-divergence - which we justify by normalizing the images and our focus on morphology, rather than intensity. In other contexts, the term I-Divergence is sometimes used for applications on data that are not (normalized) distribution functions.} as transformation information}
\label{subsec:KLdivTI}
To capture symmetry features quantitatively, we use an information-theoretic framework that measures self-similarity under geometric transformations.
Let $I(x)$ denote an image intensity on a domain $D \subset \mathbb{R}^2$, and let $I_a(x)=T_a I(x)$ be the image transformed by parameters $a$ (e.g., rotation angle or reflection axis). 
Our method builds on the Transformation Information (TI) measure of \cite{gandhi_etal_2021_TIpaper}, comparing an image $I(x)$ to its transformed counterpart $I_a(x)$ through a pixel-wise intensity ratio.
We evaluate the similarity between $I(x)$ and $I_a(x)$ on the overlap $\widetilde{D}=D\cap T_a(D)$, and define normalized intensities
\begin{equation}
p(x)=\frac{I(x)}{\int_{D} I(x)\,\mathrm{d}A}, \qquad
q_a(x)=\frac{I_a(x)}{\int_{D} I_a(x)\,\mathrm{d}A}.
\label{eq:prob-p-q}
\end{equation}
We can then write an expression for the Kullback–Leibler (KL) divergence between normalized intensity fields \citep{KullbackLibler1951} as:
\begin{equation}
D_{\mathrm{KL}}\!\left(p\,\|\,q_a\right)
= \int_{\widetilde{D}} p(x)\,\ln\!\frac{p(x)}{q_a(x)}\,\mathrm{d}A \;\;\ge 0,
\label{eq:kl}
\end{equation}
with equality if and only if $p=q_a$ almost everywhere (Gibbs’ inequality).
Consequently, \emph{symmetry detection} reduces to finding $a^\star=\arg\,\min_a D_{\mathrm{KL}}(p\|q_a)$; the minima indicate symmetry operations, and their depths quantify the symmetry strength.
Rewriting Eq.~\ref{eq:kl} in terms of the original image intensity, we arrive at the TI formalism defined in \cite{gandhi_etal_2021_TIpaper}
\begin{equation}
\mathrm{TI}(a) = \frac{1}{|\widetilde{D_\alpha}|} \int_{\widetilde{D_\alpha}} I(x)\ \ln\ \left[\frac{I(x)}{T_a I(x)}\right]\ \mathrm{d}A,
\label{eq:ti}
\end{equation}
where $T_a$ is a transformation (rotation, reflection) and $\widetilde{D_\alpha}$ denotes the overlapping domain between the image and its transform. 
We note that in eq~\ref{eq:ti}, the image and its' transform are normalized by the overlap domain, which is identical for both - hence omitted from the division between $I(x)$ and $T_a I(x)$.
Local minima of $\mathrm{TI}(a)$ correspond to transformations that preserve maximal information—i.e., symmetry axes. 
The approach is \emph{non-parametric and agnostic to geometry}, enabling object-to-object and population-level comparisons (e.g., thermonuclear supernova remnants versus CCSNRs, Section~\ref{subsection:CCSNeVStypeIa}).

A note on the asymmetric nature of KL-divergence. In our case, the KL divergence is calculated between an image, $I(x)$, and its transformed counterpart, $T_a I(x)$. While this does not eliminate the asymmetry of the KL divergence, it constrains its consequences through the transformations (here, rotations and reflections), which are invertible and measure-preserving..
Therefore, reversing the order of the KL divergence does not introduce a fundamentally different symmetry test: in the ideal continuous case, exchanging the two arguments corresponds to replacing the transformation parameter by the inverse transformation. For rotations this is equivalent to changing the sign of the rotation angle, while reflections are self-inverse.
Physically, this does not alter the inferred symmetries, since the locations of the TI minima define symmetry axes that are invariant under reversal of the divergence arguments.

\subsection{Maximum-likelihood interpretation}
The KL divergence is closely related to maximum-likelihood estimation. 
Let $x^{(1)},\dots,x^{(N)}$ be independent and identically distributed (i.i.d.) measurements drawn from an unknown data distribution $p_{\mathrm{data}}(x)$, and let $q_{\theta}(x)$ be a parametric model distribution with parameters $\theta$. 
The (population) average log-likelihood of the model is
\begin{equation}
    \mathbb{E}_{x \sim p_{\mathrm{data}}}\!\left[ \log q_{\theta}(x) \right] 
    = \int p_{\mathrm{data}}(x)\,\log q_{\theta}(x)\,\mathrm{d}x.
\end{equation}
We can rewrite this in terms of the KL divergence by adding and subtracting $\log p_{\mathrm{data}}(x)$ inside the expectation:
\begin{equation}
    \mathbb{E}_{p_{\mathrm{data}}}\!\left[ \log q_{\theta}(x) \right] 
    = \mathbb{E}_{p_{\mathrm{data}}}\!\left[ \log \frac{q_{\theta}(x)}{p_{\mathrm{data}}(x)} \right] 
\end{equation}
\vspace{-0.5cm}
\begin{equation}
\nonumber
    \hspace{\textwidth/7} + \mathbb{E}_{p_{\mathrm{data}}}\!\left[ \log p_{\mathrm{data}}(x) \right].
\end{equation}
The second term does not depend on the model parameters $\theta$, while the first term is
\begin{equation}
    \mathbb{E}_{p_{\mathrm{data}}}\!\left[ \log \frac{q_{\theta}(x)}{p_{\mathrm{data}}(x)} \right]
    = - D_{\mathrm{KL}}\!\left(p_{\mathrm{data}} \,\|\, q_{\theta}\right).
\end{equation}
Thus,
\begin{equation}
    \mathbb{E}_{p_{\mathrm{data}}}\!\left[ \log q_{\theta}(x) \right]
    = - D_{\mathrm{KL}}\!\left(p_{\mathrm{data}} \,\|\, q_{\theta}\right)
      + \mathrm{constant} \enspace (\text{w.r.t. }\theta),
\end{equation}
and maximizing the expected log-likelihood over $\theta$ is exactly equivalent to minimizing $D_{\mathrm{KL}}\!\left(p_{\mathrm{data}} \,\|\, q_{\theta}\right)$.

In our case, using Eq.~\ref{eq:prob-p-q}, the data distribution $p_{\mathrm{data}}(x)$ is identified with the normalized image $p(x)$, and the model $q_{\theta}(x)$ is the distribution obtained by applying a transformation $T_a$ to the image, i.e.\ $q_{\theta}(x) \equiv q_{a}(x)$. 
Therefore,
\begin{equation}
    \arg\max_{a}\,\mathbb{E}_{p}\!\left[ \log q_{a}(x) \right]
    = \arg\min_{a} D_{\mathrm{KL}}\!\left(p \,\|\, q_{a}\right),
\end{equation}
so the transformation parameters $a$ that minimize the KL divergence (or TI) between the original and transformed image are precisely the maximum-likelihood estimates under this self-consistency model.
We emphasize that the maximum-likelihood interpretation is merely a mathematical analogy intended for intuition, as image intensities do not constitute true probabilities.

\vspace{0.5cm}
We present our implementation of the TI measure analysis in Section~\ref{sec:Methods}, demonstrate its application to images of PNe in Section~\ref{sec:TestCalibration}, and apply the method to SNRs in Section~\ref{sec:Application}. We show-case an alternative application of the method for thresholded images of SNRs in Section~\ref{subsection:Thresholded} and draw conclusions for the differences between CCSNRs and Type Ia remnants in Section~\ref{subsection:CCSNeVStypeIa}. We summarize our results in Section~\ref{sec:Summary}.

% ================================================
\section{Methods}
\label{sec:Methods}
% ================================================

We implement the TI measure described in Eq.~\ref{eq:ti} using \textsc{Python} and apply it on a set of test images, similar to as described in \cite{gandhi_etal_2021_TIpaper}.
We first transform the image to grayscale to have a one-dimensional intensity profile for the calculation. For multi-wavelength (color) images, this `flattening' also ensures emission of different kinds is taken into account in the analysis - to gather a more complete morphological structure. Different scalings and methods of combining different kinds of emission will be tested in future papers.

For the transformations, we scan $a$ over rotations in $\theta\in[0,360^\circ)$ and reflection axes over $\theta\in[0,180^\circ)$
\footnote{Symmetry considerations; two reflection transformations separated by $180^\circ$ result in the the same reflection axis.}

After calculating the appropriate transformation matrix, we apply the transformations using either \textsc{SciPy} or \textsc{OpenCV}\footnote{We include an implementation of both for user environment compatibility. The two implementations are consistent in their output.} and calculate the resulting TI measure between the original image $I(x)$ and the transformed $T_aI(x)$, where we truncate the images to the overlap domain.
This yields $D_{\mathrm{KL}}(\theta)$ curves whose local minima give rotational and reflectional symmetries, respectively.

% FFFFFFFFFFFFFFFFFFFFFFFFFFFFFFFFFFFFFFFFFFFFFF
\begin{figure*}[htp]
    \centering
    \includegraphics[trim=0cm 0cm 0cm 0cm,width=\textwidth]{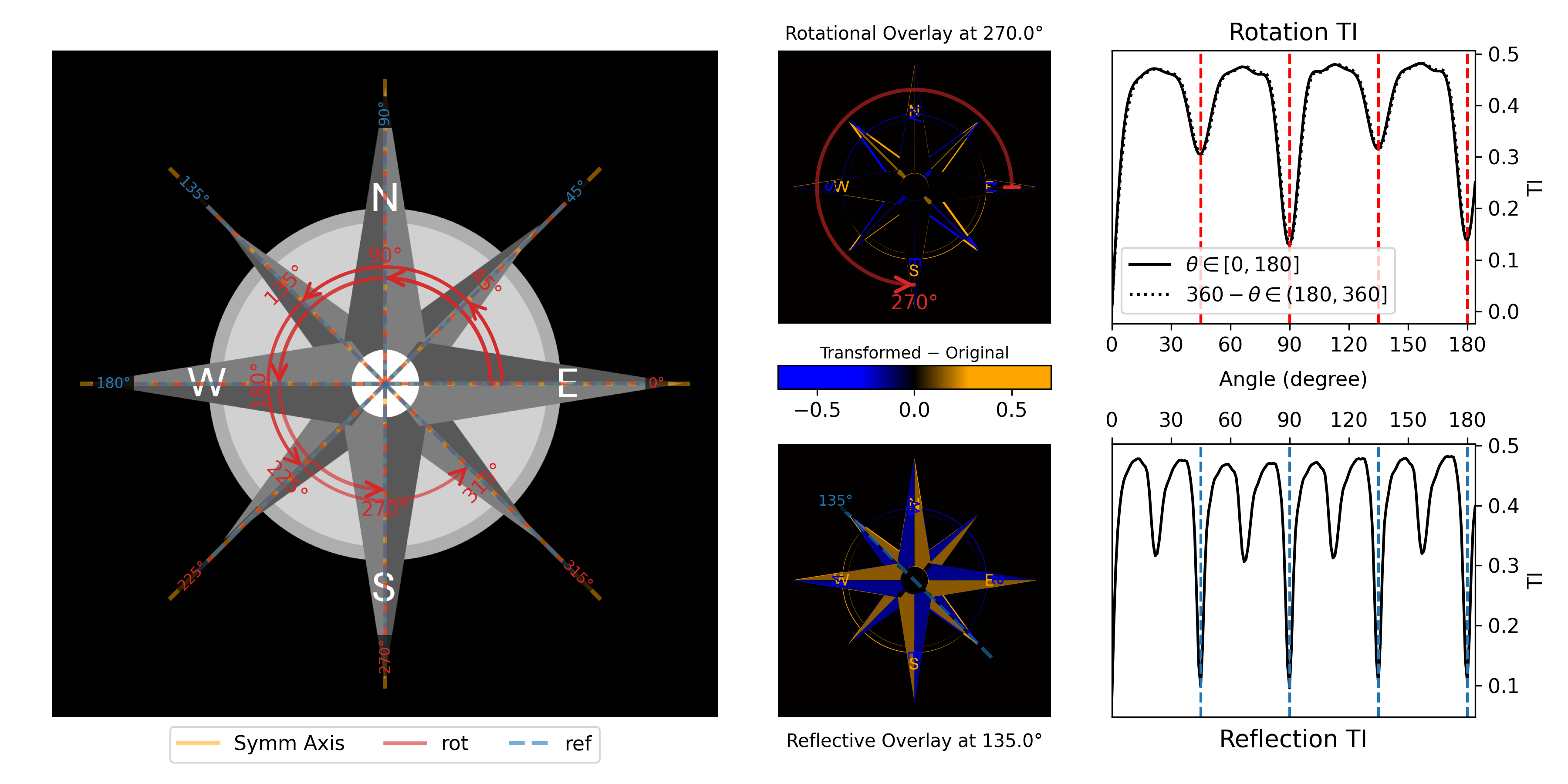} %[trim=left lower right upper]
    \caption{Our implementation of the transformation information (TI) symmetry identifier, demonstrated on a wind-rose synthetic image. Right panels depict the output of our processing pipeline. Upper right is the rotational TI measure. Dashed red vertical lines indicate identified valleys, denoting angles $\theta_{symm}^{rot}$ for which rotating the image by $\theta_{rot}$ results in a local TI minima, corresponding to a rotational symmetry.
    The bottom right panel is the reflection TI measure, with dashed blue lines indicating $\theta_{ref}$, for which a reflection about a line at $\theta_{ref}$ produces a local minima in the TI measure.
    Left is our annotated image, where we denote $\theta_{rot}$/$\theta_{ref}$ (dashed blue/red lines) on the image. The symmetry axes corresponding to $\theta_{ref}$, set at $\theta^{ref}_{symm}=\theta_{ref}+90^\circ$ are denoted as solid yellow lines, and serve to guide the eye for symmetric features along this axis. Rotational symmetries are denoted with a curved arrow, where $\theta_{symm}^{rot}$ is denoted near the arrow head. Dashed red lines indicate $\theta_{symm}^{rot}$ angles, with an additional line at $\theta=0$ for reference. 
    Two panels at the center demonstrate the subtraction between the transformed image and the original, for two examples: $\theta_{rot}=270^\circ$ (top) and $\theta_{ref}=135^\circ$ (bottom), scaled with the color bar between the two panels. 
    For our wind-rose example, all reflection symmetries matching each opposite spike pair and rotational symmetries corresponding to rotations from one spike to the next are consistent with base truth assumptions.}
    \label{fig:Example}
\end{figure*}
% FFFFFFFFFFFFFFFFFFFFFFFFFFFFFFFFFFFFFFFFFFFFFF

We use standard affine transform implementations (cv2.warpAffine and scipy.affine\_transform) with linear interpolation. 
While interpolation-induced artifacts were negligible in our analysis due to the high image resolution and focus on large-scale extended features, sub-pixel handling may become relevant in lower-resolution datasets or when studying small-scale structures.

Our full implementation is available online at a GitHub repository\footnote{\href{https://github.com/AmirGoshenMichaelis/symmetries\_snr}{symmetries\_snr} includes both the processing pipeline and examples for various example objects.}, where we also address numerical issues and limitations.

To demonstrate our implementation of the TI symmetry identifier described above, we apply it on a synthetic image featuring a wind-rose structure in Figure~\ref{fig:Example}. The image contains four large spikes at $\theta_{lp}=0^\circ,90^\circ,180^\circ,270^\circ$ and four small spikes offset by $45^\circ$ at $\theta_{sp}=45^\circ,135^\circ,225^\circ,315^\circ$. We added directional labels (N,S,W,E) to the large spikes and have purposely left the wind-rose slightly misaligned as to introduce minor imperfections (noise) and allow for differences when transformed by an even transformation. In the left panel we show the wind-rose image with overlayed symmetries, and in the right-most panels the TI measure for rotational transformations (top) and reflection transformations (bottom). In the middle panels we show an example of the original image subtracted from the rotation/reflection (top/bottom) transformed image with $\theta=180^\circ$ to illustrate the process.

Rotational TI (top right panel in Figure~\ref{fig:Example}) identifies seven rotational symmetry ($\theta_{symm}^{rot}$) regions in the range $0^\circ<\theta<360^\circ$, at $\theta_{symm}^{rot}=45^\circ,90^\circ,135^\circ,180^\circ,225^\circ,270^\circ,315^\circ$. The wider and less deep symmetries with $\Delta\theta\approx15^\circ$, at $\theta_{symm}^{rot}=45^\circ,135^\circ$ correspond to rotation causing an overlap between small spikes ($\theta_{sp}$) and large spikes ($\theta_{lp}$). The sharper symmetries at $\theta_{symm}^{rot}=90^\circ,180^\circ$, which also have a lower TI measure (and hence, are `more symmetric'), correspond to a symmetric overlap: small spikes on small spikes, large spikes on large spikes. For the rotational TI, $\theta_{symm}^{rot}=0^\circ$ corresponds to a comparison between an untransformed (un-rotated) image and the original and is hence (the only) with $\rm TI=0$.

Reflection TI (bottom right panel in Figure~\ref{fig:Example}) captures eight symmetry axes ($\theta_{ref}$), every $\delta\theta_{ij}=\theta_{symm}^i-\theta_{symm}^j\approx22.5^\circ$ in the range $0^\circ<\theta<180^\circ$. We denote four of these axes at $\theta_{symm}=\theta_{ref}+90^\circ$ as solid yellow lines on the image (left panel of Figure~\ref{fig:Example}) to illustrate their directions. These $\theta_{symm}=(45^\circ,90^\circ,135^\circ,180^\circ)+90^\circ$ represent reflection transformation that map $\theta_{sp}$ to $\theta_{sp}$ and $\theta_{lp}$ to $\theta_{sp}$ (small to small, large to large). The intermittent $\theta_{ref}=22.5^\circ,67.5^\circ,112.5^\circ,157.5^\circ$ correspond to transformations that align smaller spikes with larger spikes, resulting in a lesser match - higher TI measure.

We use these definitions of angles and a similar figure composition of denoted image on the left and TI measures (rotation and reflection) on the right for the rest of the paper.
We now turn to apply our TI analysis pipeline to images of astrophysical systems.

% ================================================
\section{PNe - Test and Calibration}
\label{sec:TestCalibration}
% ================================================
We first employ our TI symmetry analysis on planetary nebulae (PNe).  

PNe can be categorized by their dominant morphological shape. For the 250 PNe in the IAC morphological catalog \citep{1996iacm.book.....M}, $~20\%$ of the PNe are classified as either bipolar, quadrupolar or point-symmetric - having several axes of symmetry (see exact definitions and fractions in \citealt{1997IAUS..180...24M}).
\cite{Avitan_Soker_2025_PNeAngles} attempted to quantify the angles between symmetry axes, and enumerated these multiple-axis PNe to be 40, out of the over 700 PNe in the Bruce Balick planetary nebula image catalog (PNIC\footnote{Bruce Balick, PNIC: \url{https://faculty.washington.edu/balick/PNIC/}}) which includes pre-PNe. From the above, we estimate that roughly 10\% of all PNe are characterized by multiple axes.

A complimentary, more general morphological definition is "multipolar" (e.g., \citealt{Sahai_2000_M137, Wen_etal_2024_Multipolar}), i.e., having multiple bi-polar structures - which we here term symmetry axes.

\begin{figure*}[th!]
    \centering
    \includegraphics[trim=0cm 0cm 0cm 0cm,width=\textwidth]{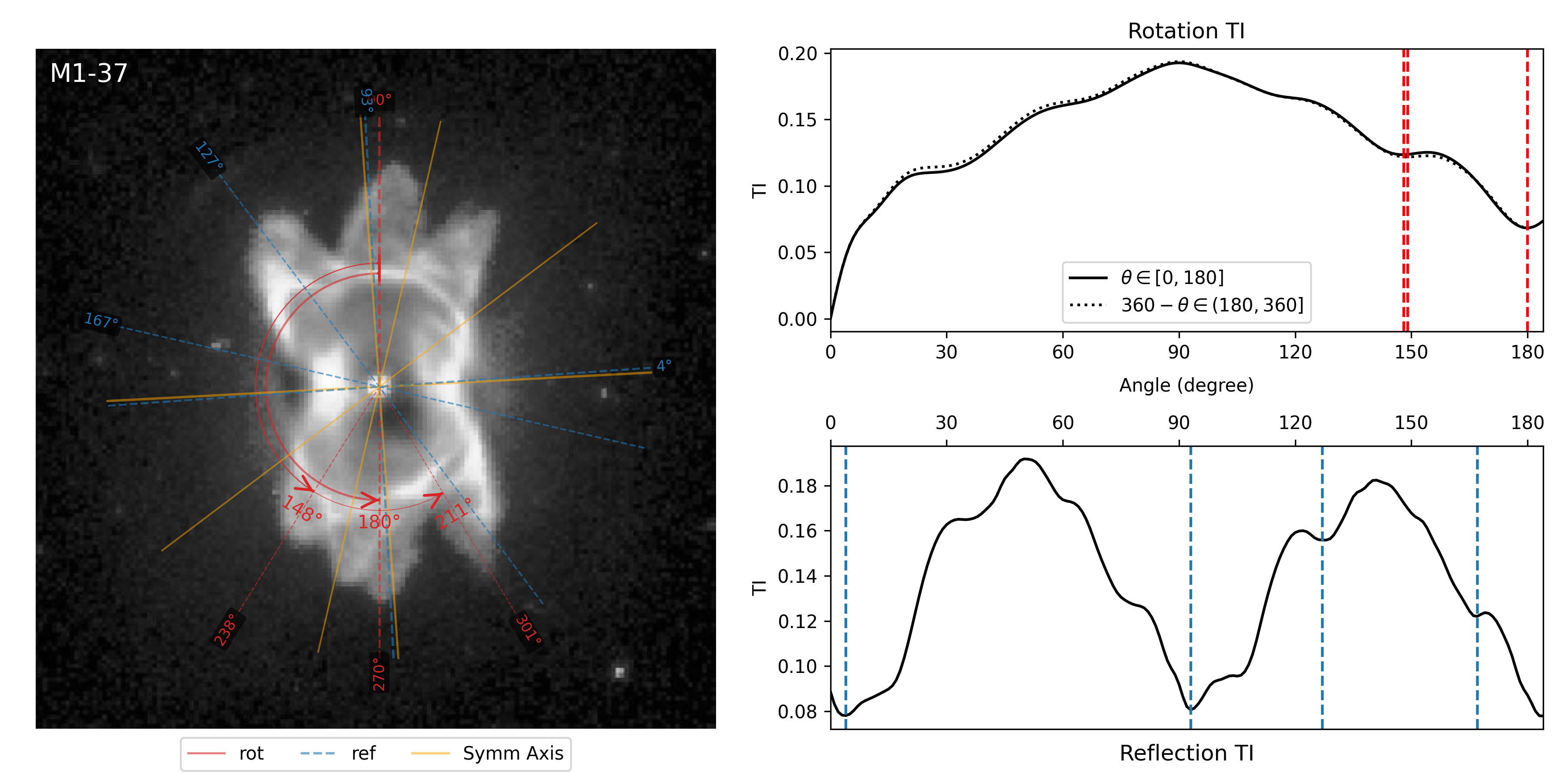} %[trim=left lower right upper]
    \caption{Our implementation of the TI symmetry identifier for the PN M1-37. Dashed-red/blue lines depict axis along which the transformations reveal the PN is symmetrical, and in solid lines the corresponding symmetry lines themselves. For the reflection symmetry the identified axes at $\theta_{symm}\approx77^\circ, 94^\circ$ degrees (solid-yellow) correspond to the (split, see lower half) vertical structure, but missing the axes at $\approx125^\circ$ and $\approx55^\circ$. An equatorial bright symmetric indent is also tagged at $\theta_{symm}=3^\circ\approx0^\circ$. Overall, the north-south symmetry is captured by a reflection axis $\theta_{ref}=4^\circ\approx0^\circ$, and a rotation axis at $\theta_{symm}^{rot}=179^\circ\approx180^\circ$. Rotation TI also recovered the missing $\theta_{symm}=55^\circ+180^\circ$.}
    \label{fig:M1-37}
\end{figure*}

In Figure~\ref{fig:M1-37} we analyze PN M1-37 \citep{Sahai_2000_M137}. M1-37 is of very prominent multi-lobed structure, with a bright center harboring the central star. The three northern 
%...
% FFFFFFFFFFFFFFFFFFFFFFFFFFFFFFFFFFFFFFFFFFFFFF
% FFFFFFFFFFFFFFFFFFFFFFFFFFFFFFFFFFFFFFFFFFFFFF
%...
(upper side of the image) poles can be assigned the following angles: $\theta_{feature}\approx55^\circ,90^\circ,125^\circ$, with opposing structures ($180^\circ$) to the south (see \citealt{Sahai_2000_M137} for notations and designations).

The TI symmetry identifier successfully recovered the symmetry axes corresponding to the main lobes: $\theta_{symm}=257^\circ, 274^\circ$ for the split lobe at $\theta_{feature}\approx270^\circ$. $\theta_{ref}=127^\circ$ corresponds to a reflection around $\theta_{feature}\approx125^\circ$. 
A rotational symmetry at $\theta_{symm}^{rot}=180^\circ$ highlights the overall symmetry of M1-37. Two secondary rotational symmetries $\theta_{symm}^{rot}=148^\circ,211^\circ$ can be interpreted as matching two lobes from the (image) upper part to two lobes in the (image) lower part. This partial matching of only two lobes (out of three) is also expressed in a higher TI (less match) of $\rm{TI}_{148^\circ}^{rot}\approx0.12$ compared to $\rm{TI}_{180^\circ}^{rot}\approx0.07$.

Overall, several symmetrical features previously identified and denoted \citep{Sahai_2000_M137} were successfully identified and quantified:
\begin{itemize}
    \item Symmetrical (reflection) lobe pair at $\theta_{feature}\approx90^\circ$ (with split at the image south)
    \item Symmetrical (reflection) at $\theta_{feature}\approx0^\circ$ (image left-right) emphasizing the bright `waist' feature.
    \item Overall $\theta_{symm}^{rot}=180^\circ$ symmetry.
    \item Rotational symmetries matching lobes from top to bottom.
\end{itemize}

\begin{figure*}
    \centering
    \includegraphics[trim=0cm 0cm 0cm 0cm,width=0.9\textwidth]{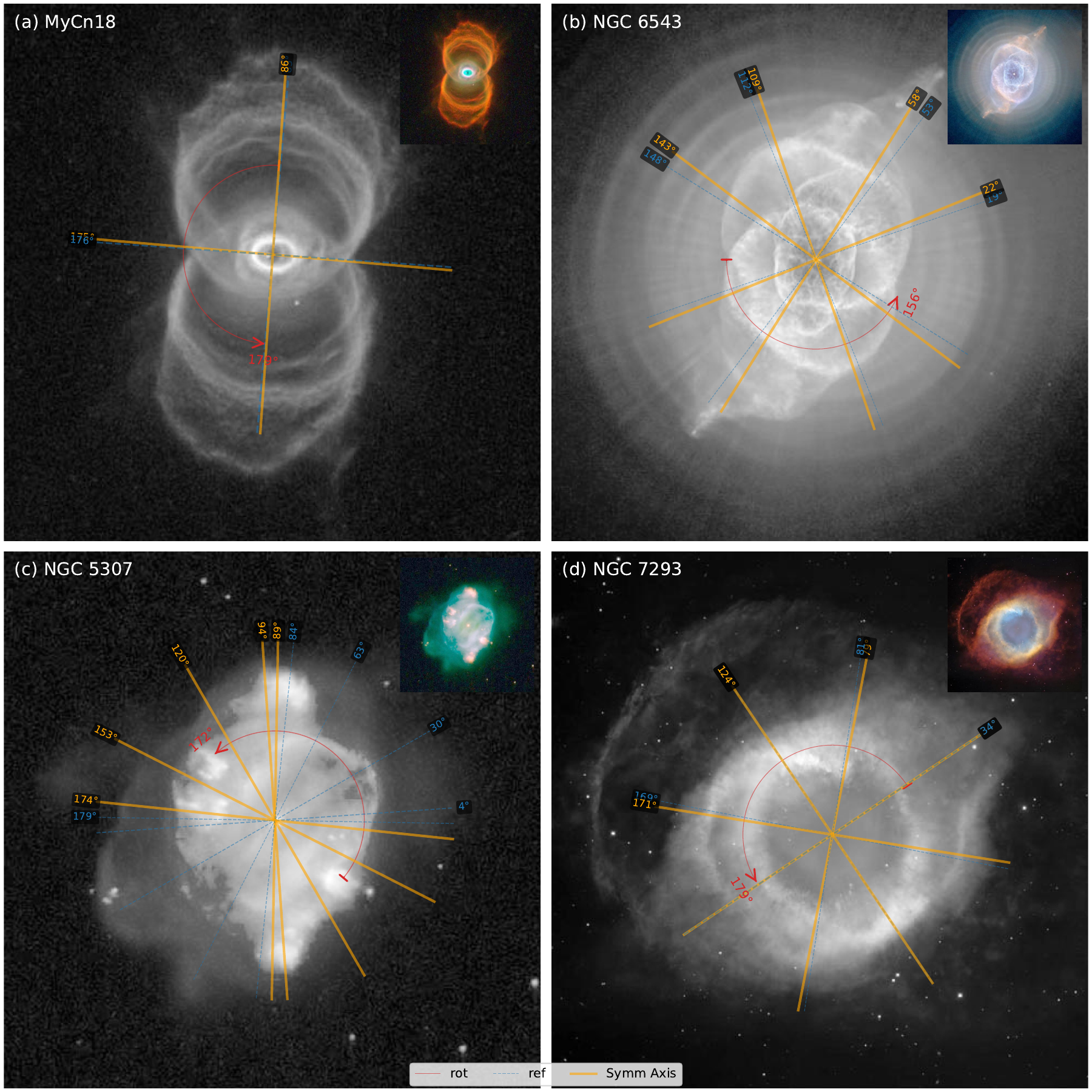} %[trim=left lower right upper]
    \caption{Four more analyzed PNe, where we denote the identified symmetry axis identified by the TI measure with yellow lines. We label the panels with their designation at the top left, and in an inset at the top right we add the original nebulae images, as taken from the Bruce Balick planetary nebulae catalog. We present the full TI profiles for these PNe in Appendix~\ref{appendix:TIplots}.\\
    \textbf{(a) MyCn 18}\footnote{The (Engraved) Hourglass Nebula. An image of this nebula was used as the album cover of the Binaural album of the Pearl Jam band.}: Two clear axis are identified, corresponding to the hourglass lobes (north-south, $\theta_{symm}=86^\circ\approx90^\circ$) and the secondary axes in the equatorial plane ($\theta_{symm}=175^\circ\approx180^\circ$). A rotational symmetry with $\theta_{symm}^{rot}=179^\circ\approx180^\circ$ corresponds to the general $180^\circ$ symmetry. \\
    % https://ui.adsabs.harvard.edu/abs/1940BHarO.913....7M/abstract
    \textbf{(b) NGC 6543}\footnote{Caldwell 6, The Cat's Eye Nebula}: four reflection symmetries and one rotational symmetry. The general elongated structure is identified by $\theta_{symm}=58^\circ$. Large features corresponding to the complicated ringed structure are matched by both $\theta_{symm}=22^\circ, 109^\circ, 143^\circ$ and the rotational symmetry of $\theta_{symm}^{rot}=156^\circ$. The many concentric circular shells (e.g., \citealt{Guerrero2020_NGC6543rings}) did not interfere with the TI measure as these are approximately uniformly distributed along all angles. 
    \\
    \textbf{(c) NGC 5307}: The main north-south elongated structure featuring two opposite bright regions is identified by the $\theta_{symm}=89^\circ, 94^\circ$ axes. The bright knots opposite at $\theta_{feature}\approx150^\circ$ triggered an identified axes at $\theta_{symm}=153^\circ$ and potentially with the two north-south knots, $\theta_{symm}=174^\circ$. A rotational symmetry axis at $\theta_{symm}^{rot}=172^\circ\approx180^\circ$ is likely a combination of the bright knots and the general polar shape. \\
    \textbf{(d) NGC 7293}\footnote[4]{Caldwell 63, The Helix Nebula}: Four reflection axes are detected, roughly $45^\circ$ from one another. $\theta_{symm}=34^\circ$ appears to coincide with the general elongated shape, and the other axes at $\theta_{symm}=79^\circ,124^\circ,171^\circ$ with bright rims that are equidistant and located opposite from the center. A rotational symmetry axis $\theta_{symm}^{rot}=179^\circ$ corresponds to a bright bipolar shape.
    }
    \label{fig:PNe}
\end{figure*}

As to further test the method for consistency in feature identification and tagging, we now apply the TI symmetry identifier to a set of four additional PNe. In Figure~\ref{fig:PNe} we present the annotated output of our method, similar to the left panel of Figure \ref{fig:M1-37}.
% FFFFFFFFFFFFFFFFFFFFFFFFFFFFFFFFFFFFFFFFFFFFFF
% FFFFFFFFFFFFFFFFFFFFFFFFFFFFFFFFFFFFFFFFFFFFFF

For all four PNe tested in Figure~\ref{fig:PNe}, strong symmetrical features were successfully identified and tagged:
\begin{itemize}
    \item MyCn 18: A $\theta_{symm}^{rot}=179^\circ\approx180^\circ$ rotational symmetry highlights the bipolar structure. Additionally, two axes correspond to the bright rims ($\theta_{symm}=86^\circ\approx90^\circ$, image top-down, e.g., \citealt{Sahai_etal1999_MYCn18_obs}) and the extended hourglass central waist ($\theta_{symm}=175^\circ\approx180^\circ$, image left-right).
    \item NGC 6543: The outer bubble-like structures are pointed at with $\theta_{symm}=109^\circ$ and other more inner structures (e.g., \citealt{BalickHajian2004_NGC6543_Features}) have triggered $\theta_{symm}=22^\circ,143^\circ$. The rotational symmetry of $\theta_{symm}^{rot}=156^\circ$ appears to also correlate with the bright edges of two such bubbles. The extended jets are tagged by $\theta_{symm}=58^\circ$.
    \item NGC 5307: Five reflection symmetry axes are identified, corresponding to features along $\theta_{feature}\approx90^\circ$ ($\theta_{symm}=89^\circ,94^\circ$) and $\theta_{feature}\approx150^\circ$ ($\theta_{symm}=153^\circ,174^\circ$). These appear to be correlated with the point symmetric outflows of NGC 5007 (e.g., \citealt{Bond_etal1995_NGC5007_PointSymmetric}). A rotational symmetry of $\theta_{symm}^{rot}=172^\circ$ seems to match the same point symmetric features.
    \item NGC 7293: Similar to NGC 6543 - symmetry axes ($\theta_{symm}=79^\circ,124^\circ,171^\circ$) correspond to opposite large scale equidistant symmetrical features. An additional axis at $\theta_{symm}=34^\circ$ denotes the two northwest-southeast outer features (e.g., \citealt{ODell_etal2004_NGC7293_features}) and $\theta_{symm}^{rot}=179^\circ\approx180^\circ$ emphasizes again this symmetry axis.
\end{itemize}
In cases where point symmetrical features were extended or part of structures not along a line originating from the center, the symmetry axes we denote with yellow lines only denote the vertical to the transformation axis; E.g., $\theta_{symm}=174^\circ$ in NGC 5307, where a bright outflow is visible at $\theta_{feature}=173^\circ$ but not at $\theta=173^\circ-180^\circ=-6^\circ$.

% ================================================
\section{Application for SNR\texorpdfstring{\MakeLowercase{s}}{s}}
\label{sec:Application}
% ================================================

While it is widely accepted that many PNe have a structured bi-polar or point symmetric structure (see Section~\ref{sec:TestCalibration}), SNRs, and specifically core-collapse SNRs (CCSNRs), are much less organized in their remnant morphology.

\begin{figure*}[ht!]
    \centering
    \includegraphics[trim=0.7cm 0cm 0cm 0cm,clip,width=1.02\textwidth]{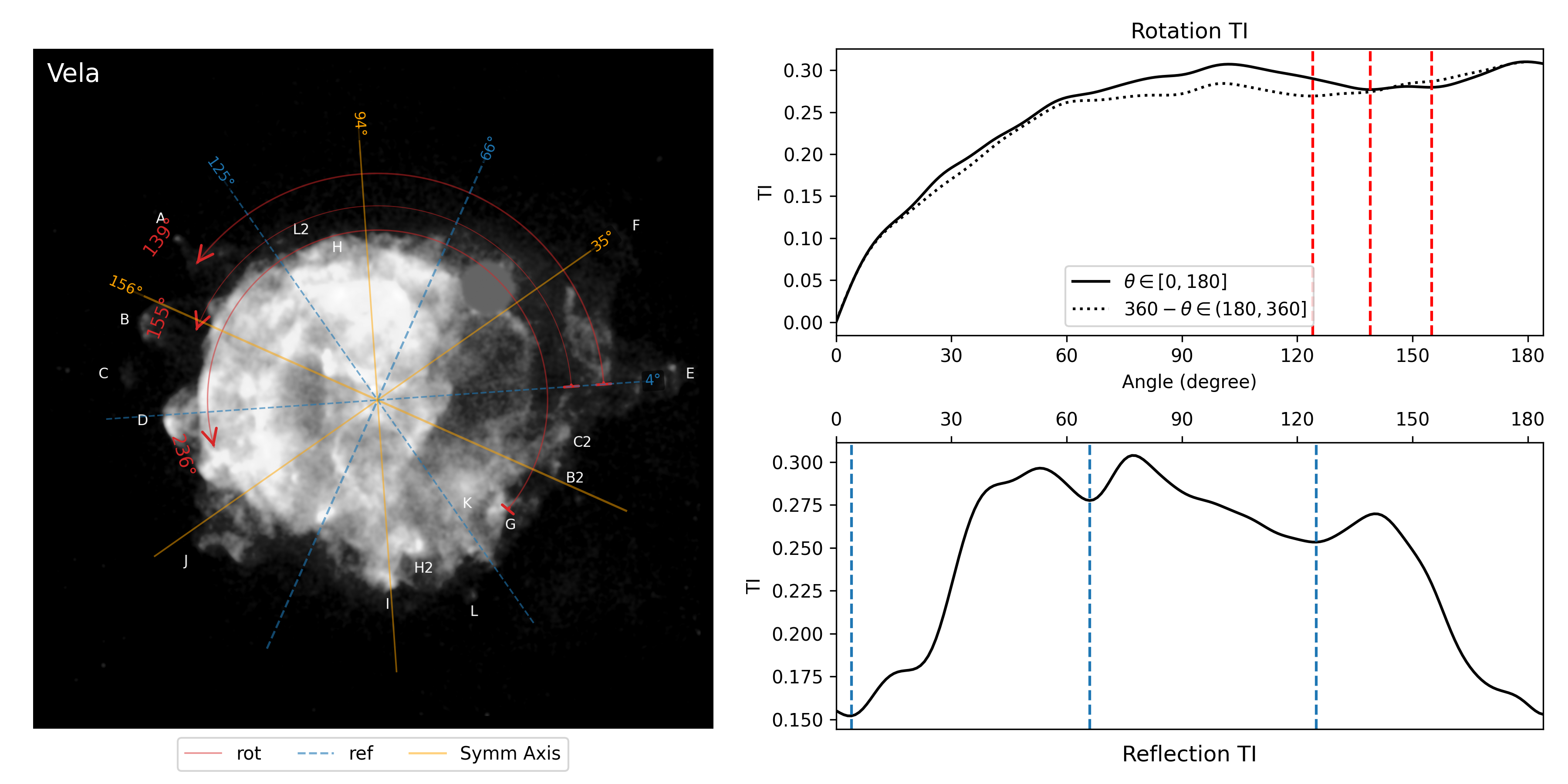} %[trim=left lower right upper]
    \caption{Application of the TI measure on a soft X-ray image of the Vela SNR (eROSITA DR1 data, see details and references in \citealt{SokerShishkin2025_Vela}). To separate the Vela SNR from two other SNRs in the same patch of sky, we excise the Puppis A SNR from the image replacing a circle around its' center with the average of its surrounding, and use an image of the lower energy spectrum where the Vela Jr. SNR is less visible.
    % We chose the center location to be the current location of the Vela pulsar.
    We chose the center location to be the deduced center between opposite clumps (blue star in Figure 1 of \citealt{SokerShishkin2025_Vela}).
    Several projectiles (clumps) and their bow-shock wake are accompanied by bright X-ray emission. We denote previously identified clumps with their associated designated letters as overlay (post-TI analysis).
    Two symmetry axes along the horizontal image direction $\theta_{symm}=-24^\circ,35^\circ$ correspond to the presence of several projectiles and the slightly extended structure in this general direction.
    A symmetry axis at $\theta_{symm}=94^\circ$ correlates to a previously identified north-south axis \citep{Mayer_etal_2023_Vela, SokerShishkin2025_Vela}, and emphasizes the elongated shape extending between features D and E.
    Rotational symmetries at $\theta_{symm}^{rot}=139^\circ,155^\circ$ appear to correspond to aligning projectile track features in the outer structure of the remnant. We illustrate this with curved concentric arrows between E and A,B.
    }
    \label{fig:Vela}
\end{figure*}

We first test our method on the Vela CCSNR, noted for many ejecta knots and protruding features extending beyond the main shell (e.g., \citealt{SokerShishkin2025_Vela}).
In Figure~\ref{fig:Vela} we apply the TI measure on an X-ray image of the Vela SNR, modified to both reveal subtle emission features (with log scaling of smoothed counts) and remove other SNRs that occupy the same sky region (with masking and the selection of a soft range of X-ray emission).
% FFFFFFFFFFFFFFFFFFFFFFFFFFFFFFFFFFFFFFFFFFFFFF
% FFFFFFFFFFFFFFFFFFFFFFFFFFFFFFFFFFFFFFFFFFFFFF

Rotational TI resulted in several rotational symmetries; $\theta_{symm}^{rot}=139^\circ$ coinciding with a rotation overlapping features E and A (and to a lesser extent F with D), and $\theta_{symm}^{rot}=155^\circ$ overlapping features E and B. $\theta_{symm}^{rot}=236^\circ$ appears to match bright inner features. Rotational transformations where features along the outskirts of the remnant are matched can reveal subtle symmetries, but are sensitive to the image exposure (intensity) and scaling (See also \ref{subsection:Thresholded}).
The two symmetry axes residing approximately along the image left-right direction ($\theta_{symm}=-24^\circ,35^\circ$), do not match exactly the pairs in these directions: F-J (along $\theta_{FJ}=35^\circ$) and B-B2 ($\theta_{BB2}=-24^\circ$). Slight variations in the selection of the center changed the exact direction and number of symmetry axes identified by the TI method (not shown), but the trend of a single vertical axis and several horizontally oriented axes remained. The dependence of the TI measure scores and observed symmetries on the center choice (See also \citealt{gandhi_etal_2021_TIpaper}) is the subject of a future study.

\begin{figure*}
    \centering
    \includegraphics[trim=0cm 0cm 0cm 0cm,width=0.9\textwidth]{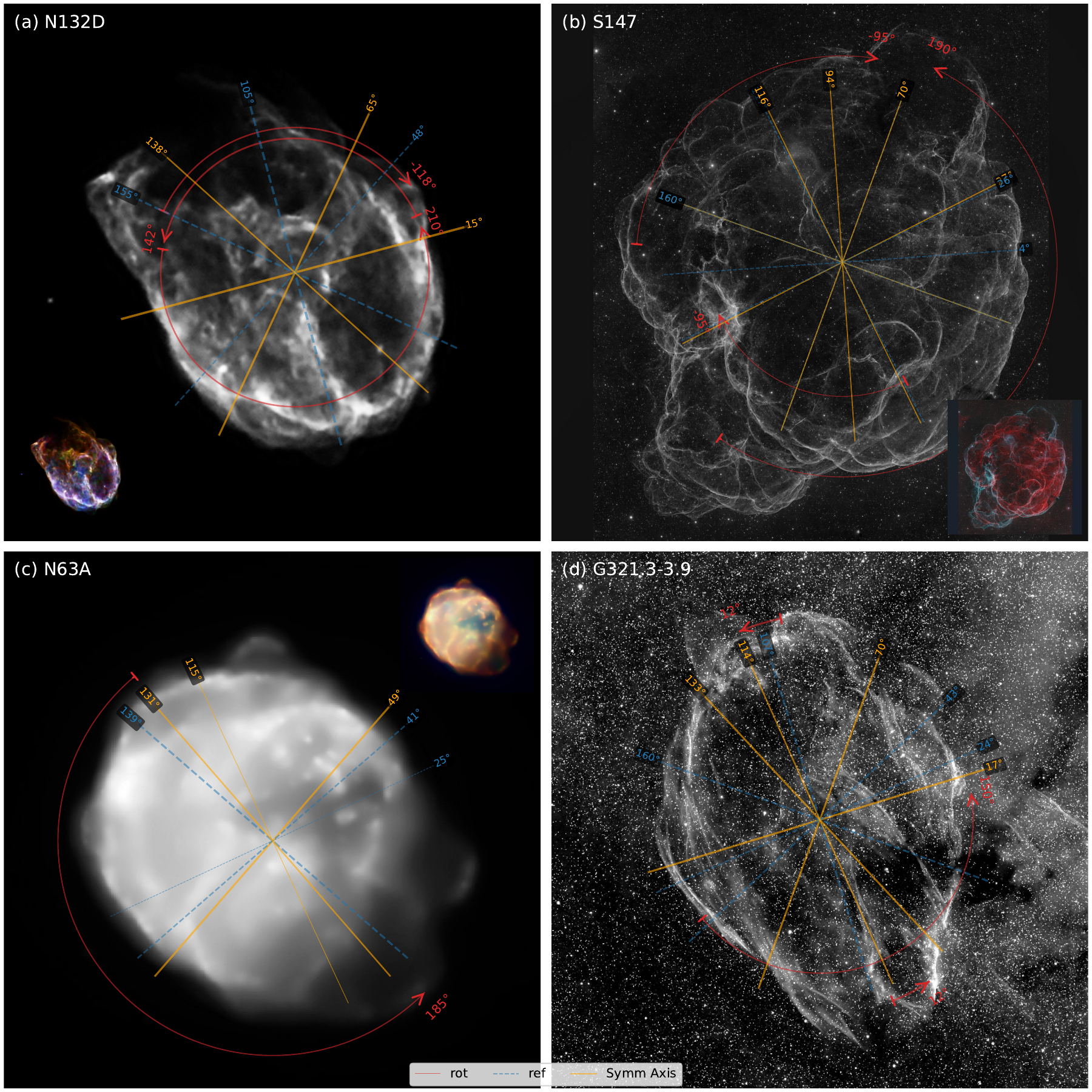} %[trim=left lower right upper]
    \caption{As Figure~\ref{fig:PNe}, but for four SNRs. \\
    \textbf{(a) N132D}\footnote{Chandra ACIS X-ray image, NASA/CXC/NCSU/K.J.Borkowski et al.}: A reflection symmetry axis at $\theta_{symm}=138^\circ$ corresponds to the extended structure (image top-left to bottom-right). Two additional symmetry axes at $\theta_{symm}=15^\circ,65^\circ$ match different parts of the bright arc at the remnant bottom, or folding along the image top-left ($\theta_{ref}=155^\circ$). Two rotation symmetries correlate with moving outer protrusions (indicated with red arrows). \\
    \textbf{(b) Simeis 147}\footnote{The Spaghetti Remnant. Image by \href{https://app.astrobin.com/i/vh6kx6}{Mr. Christian Koll}, see \citealt{ShishkinBearSoker_S147} for details.}: Several symmetry axes are identified, two coincide with the extended shape (at $\theta_{symm}=70^\circ,94^\circ$) and three ($\theta_{symm}=27^\circ,116^\circ,160^\circ$) correlate with folded-over bright zones along the outer shape. A rotational symmetry with $\theta_{symm}^{rot}=265^\circ=-95^\circ\approx90^\circ$ seems to match bright features or protrusions in the outer shape, and $\theta_{symm}^{rot}=190^\circ\approx180^\circ$ match the overall extended shape. \\
    \textbf{(c) N63A}\footnote{Chandra ACIS X-ray image, NASA/CXC/Rutgers/J.Warren et al.}: Two groups of reflection symmetries are identified; $\theta_{symm}=131^\circ$, corresponding to folding over the elongated axis (top left to bottom right in the image). Notable features coinciding over this axis are the two protruding features on both ends of the remnant.
    $\theta_{symm}=49^\circ$, similarly folds over the narrow direction, reflecting around the large tail at the image bottom right and matching the two smaller protrusions at the image top left (Ear 1 and Ear 2, see \citealt{Karagoz_etal2023_N63A} for their notations).
    A rotational symmetry of $\theta_{symm}^{rot}=185^\circ\approx180^\circ$ further cements a symmetrical opposing morphological structure. \\
    \textbf{(d) G321.3-3.9}\footnote{Visible [O III] image by Mathew Ludgate from \citealt{Fesen_etal2024_G321Optical}.}:  Of the four reflective symmetry axes, two axes $\theta_{symm}=114^\circ,133^\circ$ coincide with two structures: a protrusion on the image top left and a rimmed bubble at the image bottom right. A rotational symmetry of $\theta_{symm}^{rot}=12^\circ$ matches both the width of said bubble and the distance to a structure adjacent to the top protrusion. Bright rim-like features on the remnant narrow section (horizontal, at $\theta_{axis}\approx20^\circ$) are matched with both $\theta_{symm}=17^\circ$ and $\theta_{symm}^{rot}=150^\circ$.
    }
    \label{fig:SNRs}
\end{figure*}

We apply our symmetry identifier on a set of four additional SNRs: N132D, S147, N63A and G321.3-3.9. We chose these remnants for their complicated structure, filamentary features, incomplete shells, multiple protrusions and the availability of high resolution images. For S147 and G321.3-3.9, we use images in the visible, whereas for N63A and N132D we use X-ray images.
In Figure~\ref{fig:SNRs} we show the annotated images of these four SNRs in a format similar to FIgure~\ref{fig:PNe}.
% FFFFFFFFFFFFFFFFFFFFFFFFFFFFFFFFFFFFFFFFFFFFFF
% FFFFFFFFFFFFFFFFFFFFFFFFFFFFFFFFFFFFFFFFFFFFFF

For all four SNRs in Figure~\ref{fig:SNRs} symmetry axes identified by the TI method could be easily associated with features in the remnant structure: protrusions, bright rims, or regions with similar sub structures or bright emission (high pixel intensity).
\begin{itemize}
    \item N132D: $\theta_{ref}=48^\circ$ correlates with the extended axis (along $\theta_{symm}=138^\circ$), and $\theta_{ref}=105^\circ,155^\circ$ seem to match folding over parts of the bright outer structure at the remnant bottom right. Two pairs of complementary (summing up to $360^\circ$) rotational symmetries seem to correspond to matching brighter protrusions at the remnant image upper half (Figure~\ref{fig:SNRs}a).
    \item S147: Rotational symmetries seem to accurately capture folding bright protrusions over their counterparts as illustrated in \ref{fig:SNRs}b $\theta_{symm}^{rot}=190^\circ,255^\circ$. The $\theta_{symm}^{rot}=255^\circ=-95^\circ$ also seems to match rotating a bright inner filamentary structure over its' counterpart - possibly explaining why this rotational symmetry is most prominent (See Appendix~\ref{appendix:TIplots}).
    \item N63A: $\theta_{symm}^{rot}=185^\circ$ attests to the general polar configuration of the remnant. $\theta_{symm}=115^\circ,131^\circ$ extend the long axis of the remnant, possibly linking the ear-like protrusion in the image upper left with a similar shape at the image bottom right. We discuss the results of the TI analysis of this remnant, its' matching to previously discussed features \citep{Karagoz_etal2023_N63A} and a modification of our algorithm to focus on the outer shape in Section~\ref{subsection:Thresholded}.
    \item G321.3-3.9: Two reflection symmetries coincide with an extended ear-like feature extending along the $\theta_{feature}\approx120^\circ$ diagonal: $\theta_{symm}=114^\circ,133^\circ$. A rotational symmetry of $\theta_{symm}^{rot}=12^\circ$ matches both the width and the distance to an adjacent similar structure (along the $\theta_{feature}\approx95^\circ$ axis). Two other reflection symmetries at $\theta_{symm}=107^\circ,160^\circ$ correlate with folding over bright edge features, as does a rotational symmetry of $\theta_{symm}^{rot}=150^\circ$.
\end{itemize}

Applying the TI symmetry identifier tool to SNRs produces mixed results. While multiple symmetry axes are systematically identified, it is difficult to claim a quantitative result without previously established ground-truth symmetry axes. For remnants where some features have been previously tagged and debated we see partial success, but see Section~\ref{subsection:Thresholded} for an additional take on N63A and the features of N63A in \cite{Karagoz_etal2023_N63A} and \cite{Soker2024_N63A}.
Visible images of remnants exhibit a more complicated and less extended sub-structure as the source of emission is mainly from interacting filaments. 
Despite this, for the two remnants we tested (S147 in \ref{fig:SNRs}b and G321.3-3.9 in \ref{fig:SNRs}d) prominent features were successfully identified, attesting that this filamentary structure is not a strong setback to the TI method.

A systematic analysis using the TI symmetry identifier method can produce a quantitative parameter for remnant comparison. We propose one such application, using the prominence and width of identified peaks\footnote{We use the terminology ``peak'' when referring to local minima in the TI scores, as the identification was performed on negated profiles.} in the TI score (See Section~\ref{subsection:Thresholded}). 
We compare the TI scores produced from images of CCSNRs and Type Ia SNRs in Section~\ref{subsection:CCSNeVStypeIa} and find that these can be presented as separate populations.

% =========================================
\subsection{Two-value threshold variant of the TI symmetry pipeline}
\label{subsection:Thresholded}
% =========================================

As an additional adaptation to the processing pipeline, we apply the TI symmetry analysis to a binarized (thresholded) version of the image, whose pixel values are representation of the average value of remnant main region and the background value. As in the main analysis, we compare the image to a rotated or reflected version of itself.
This section therefore differs from the procedure used throughout the paper only in the use of a thresholded binary image, rather than the original (continuous-valued) intensity map.

\begin{figure*}
    \centering
    \includegraphics[trim=4cm 3.4cm 2cm 4.35cm,clip,width=1\textwidth]{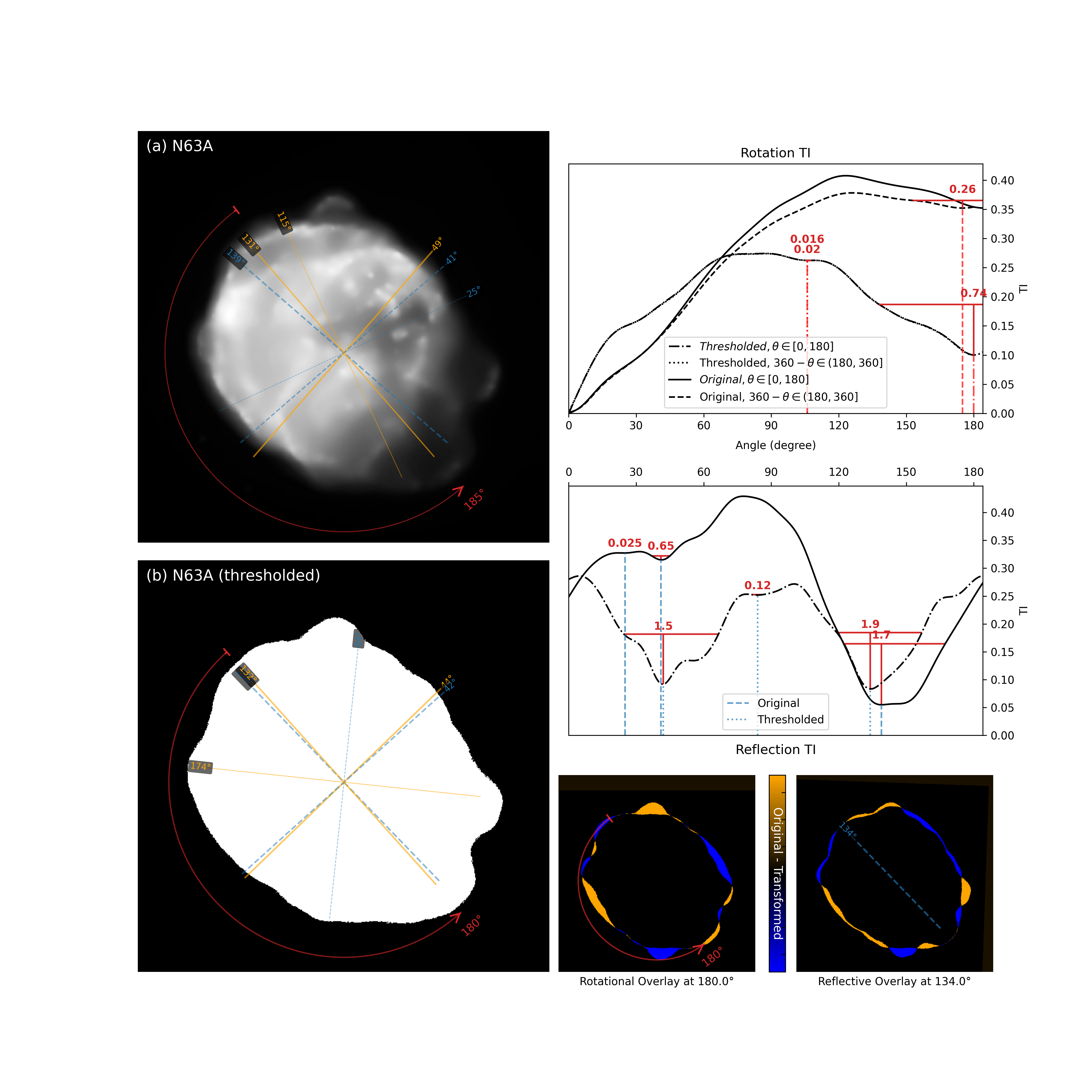} %[trim=left lower right upper]
    \caption{Demonstration of the TI symmetry identifier on a thresholded version of the SNR N63A image. \\
    On the left we show two annotated version of the N63A SNR:
    \textbf{(a)} Original image, as in Figure~\ref{fig:SNRs} panel (c). We denote identified symmetry axes ($\theta_{symm}=49^\circ, 115^\circ, 131^\circ$), and $\theta_{symm}^{rot}=185^\circ$.
    \textbf{(b)} Thresholded image. The two most prominent reflection symmetries ($\theta_{symm}=44^\circ, 132^\circ$) and a single rotation symmetry are nearly identical to the original image axes, emphasizing their strength. The third symmetry axis, $\theta_{symm}=174^\circ$, is new and correlates with a pair of protrusions not tagged by the original pipeline.
    On the right we show the rotation and reflection TI measures (top and middle panels), where solid/dashed lines denote the original image TI measure and identified symmetry axes, and dotted-dashed/dotted lines denote the thresholded version respectively.
    Note the upper right rotation TI panel has four identified symmetry axes, two of which (at $\approx105^\circ$) are overlapping.
    Bottom right two panels illustrate one rotational (left) and one reflective (right) transformations, where the untransformed image is shown in orange and the transformed in blue. The reflective symmetry about $\theta_{ref}=134^\circ$ ($\theta_{symm}=44^\circ$) matches very clearly, as shown with the mostly void subtraction image validating intuitively the TI score identification. The rotation symmetry of $\theta_{symm}^{rot}=180^\circ$ matches emphasizes the overall $180^\circ$ symmetry of the thresholded version, as also in the original image.
    }
    \label{fig:AppA_thresh}
\end{figure*}

Starting from the image $I(\mathbf{x})$, we construct a 2-value map
\begin{equation}
B(\mathbf{x}) \equiv 
\begin{cases}
  \tau &I(\mathbf{x}) \leq \tau   \\
  \,\overline{I(\mathbf{x})}\,  &I(\mathbf{x}) > \tau  
\end{cases}
\end{equation}
where $\tau$ is a chosen threshold and $\overline{I(\mathbf{x})}$ is the mean of the remnant main region. 
For the example shown in Figure~\ref{fig:AppA_thresh}, we set $\tau$ to eight times the mean intensity measured in an outer frame of width $1/20$ image size worth of pixels. Both the multiplicative factor and the frame width were chosen empirically for this demonstration, with the goal of producing a clean binary outline of the remnant. A more general application of this variant would require a self-consistent thresholding prescription, designed specifically to enhance edge and silhouette features.

To search for candidate symmetries, we apply the same family of geometric transformations as in the main pipeline:
(i) rotations by an angle $\theta$ about the adopted center, and (ii) reflections about an axis at angle $\theta$ through the same center.
Because the 2-value map isolates the object morphology, it can also be used to define an alternative center estimate (e.g., via the center of mass of $B$), which may better reflect an intuitive geometric ``center'' of the object.
For each $\theta$, we compute the TI score between the 2-value image and its transformed counterpart, producing TI curves for rotation and reflection, as shown in Figure~\ref{fig:AppA_thresh} for the SNR N63A.
% FFFFFFFFFFFFFFFFFFFFFFFFFFFFFFFFFFFFFFFFFFFFFF
% FFFFFFFFFFFFFFFFFFFFFFFFFFFFFFFFFFFFFFFFFFFFFF

The thresholded processing variant demonstrated in Figure~\ref{fig:AppA_thresh} was successful in identifying a new distinct symmetry axis (compared to the regular image analysis), that correlates to connecting a pair of previously established prominent features of the N63A remnant along the image upper left to lower right diagonal (Ear 3 and Tail 2,3 in \citealt{Karagoz_etal2023_N63A} notations).

Interpreting this variant is straightforward: because the input image takes only two intensity levels, the TI score is controlled mainly by how well the two-level morphology of the remnant matches under a transformation. In practice, the score is therefore driven primarily by the degree of pixelwise mismatch between $B(\mathbf{x})$ and its transformed counterpart $B_a(\mathbf{x}) \equiv T_a B(\mathbf{x})$.
A complementary way to visualize the symmetry quality is through a residual map such as $B(\mathbf{x})-B_a(\mathbf{x})$:
regions of near-cancellation indicate good alignment, whereas coherent residual structures indicate systematic mismatch.
The two bottom-right panels of Figure~\ref{fig:AppA_thresh} illustrate this behavior by overlaying the transformed and untransformed two-level silhouettes.

Because the map is restricted to two levels, the residuals tend to be more spatially coherent than in the continuous-intensity case: small differences in surface brightness within the remnant are suppressed, while departures in the outline and low-surface-brightness extensions become more prominent. As a result, the thresholded TI analysis emphasizes global morphology (and edge features) at the expense of internal structure, providing a useful complement to the intensity-based TI measure used in the main pipeline.

Using eq.\ref{eq:ti} with the 2-value representation of the image we arrive at the following:
\begin{equation}
\mathrm{TI}_B(a) = \frac{1}{|\widetilde{D_\alpha}|} \int_{\widetilde{D_\alpha}} B(x)\ \ln\ \left[\frac{B(x)}{T_a B(x)}\right]\ \mathrm{d}A \\
\end{equation}
\begin{equation}
\nonumber
=
\begin{cases}
  0 &B(\mathbf{x}) = B_a(\mathbf{x})   \\
  \frac{A_{\text{mr}}}{|\widetilde{D_\alpha}|}\left(\overline{I(\mathbf{x})}-\tau\right)\ln\left[\frac{\overline{I(\mathbf{x})}}{\tau}\right] &B(\mathbf{x})\neq B_a(\mathbf{x}) 
\end{cases}
\label{eq:tiB}
\end{equation}
where $A_{\text{mr}}$ is the area disagreement between $B(\mathbf{x})$ and $B_a(\mathbf{x})$ as described above (mismatched regions).

From the TI measure ($TI(a)=TI(\theta)$) we extract symmetry axes by looking for local minima, as explained in Section~\ref{subsec:KLdivTI}. We define a minima grading score based on the detected minima (\texttt{scipy.signal.find\_peaks}),
\begin{equation}\label{eq:minima_grading_score}
R_p=\frac{\text{minima prominence}}{\text{minima width}}    
\end{equation}
the ratio between the minima prominence (depth) and minima width.
We denote the minima width and prominence as calculated with \texttt{scipy.signal.peak\_widths} and \texttt{scipy.signal.peak\_prominences} in the right panels of Figure~\ref{fig:AppA_thresh}. Widths and prominences are marked with solid red lines, and the resulting minima score ($R_p$) with red text above the minima.

Overall, we find that the binarized and intensity-based pipelines recover similar symmetry angles for this example, with the thresholded variant detecting an additional symmetry axis stemming from placing a greater emphasis on the outer morphology.

The outer morphology of SNRs can reveal features and protrusions that might contain information about asymmetries stemming from the explosion mechanism or the CSM, both of which are detrimental to understanding SNe \citep[e.g.,][]{Soker2025_SNIP0509,Das_etal2025_TypeIaExp,Ravi2026_CaST}.
We therefore see the thresholded pipeline a complimentary method to the intensity-based method used throughout the paper, highlighting morphological features with lesser (pixel) intensity and potentially revealing additional symmetry axes.

% =========================================
\subsection{CCSNe vs Type Ia}
\label{subsection:CCSNeVStypeIa}
% =========================================

It has been long assumed and subsequently shown that the remnants of core-collapse and Type Ia SNe have quantitative differences, stemming from their different formation channels (e.g., \citealt{Lopez_etal2009_SNRMorph, Lopez_etal2011_SNRMorph, Peters_etal2013_SNRMorph}).
However, several recent works have attempted to upend this assumption, using novel methods or improved systematics to show the general population of SNRs cannot be subdivided into CCSNRs and Type Ia remnants according to their morphological features (e.g., \citealt{RanasingheLeahy2019_SNRMorph, Leahy_etal_2025_MultipleRemnants}).

We turn to our TI method to produce a quantitative measure of symmetry based on the detected peaks and their properties. In Section~\ref{subsection:Thresholded} we demonstrated how the TI profile peaks can be quantified based on two parameters: peak prominence and peak width.
The peak prominence can be seen as a measure of how significant the similarity between the transformed image and the original - and hence impactfull on the TI score. Where it is large, the `overlap' is large, and where small, the overlap is minimal.
The peak width can imply how localized the similar features are in $\theta$ transformation space. A narrow peak (low width) would mean features that are only overlapping for transformations in a narrow range of angles. Large width would suggest this feature extends over a wider region.

In SNRs, many thermonuclear origin remnants (Type Ia) exhibit a prominent circular morphology (e.g., \citealt{Vink2012_revSNRXray, Peters_etal2013_SNRMorph}), with slight protrusions that are wide compared to the remnant side (e.g., \citealt{Soker2024review}). This overall spherical shape and large, spread structures would result in TI measure profile that is low, with peaks that are wide. For any arbitrary transformation the resulting TI score will be low (good overlap) and a spread feature that will produce a local minima in the TI will produce a wide yet shallow (low prominence) peak.
Contrarily, in CCSNRs, the inner and outer structures tend to have a less standardizable formation (e.g., \citealt{WangWheeler2008_SNeAsym}). As a result, the overall TI measure profile will be generally high, and TI minima will be localized to narrow ranges where features overlap by the transformations.

\begin{figure*}
    \centering
    \includegraphics[trim=0cm 0cm 0cm 0cm,width=\textwidth]{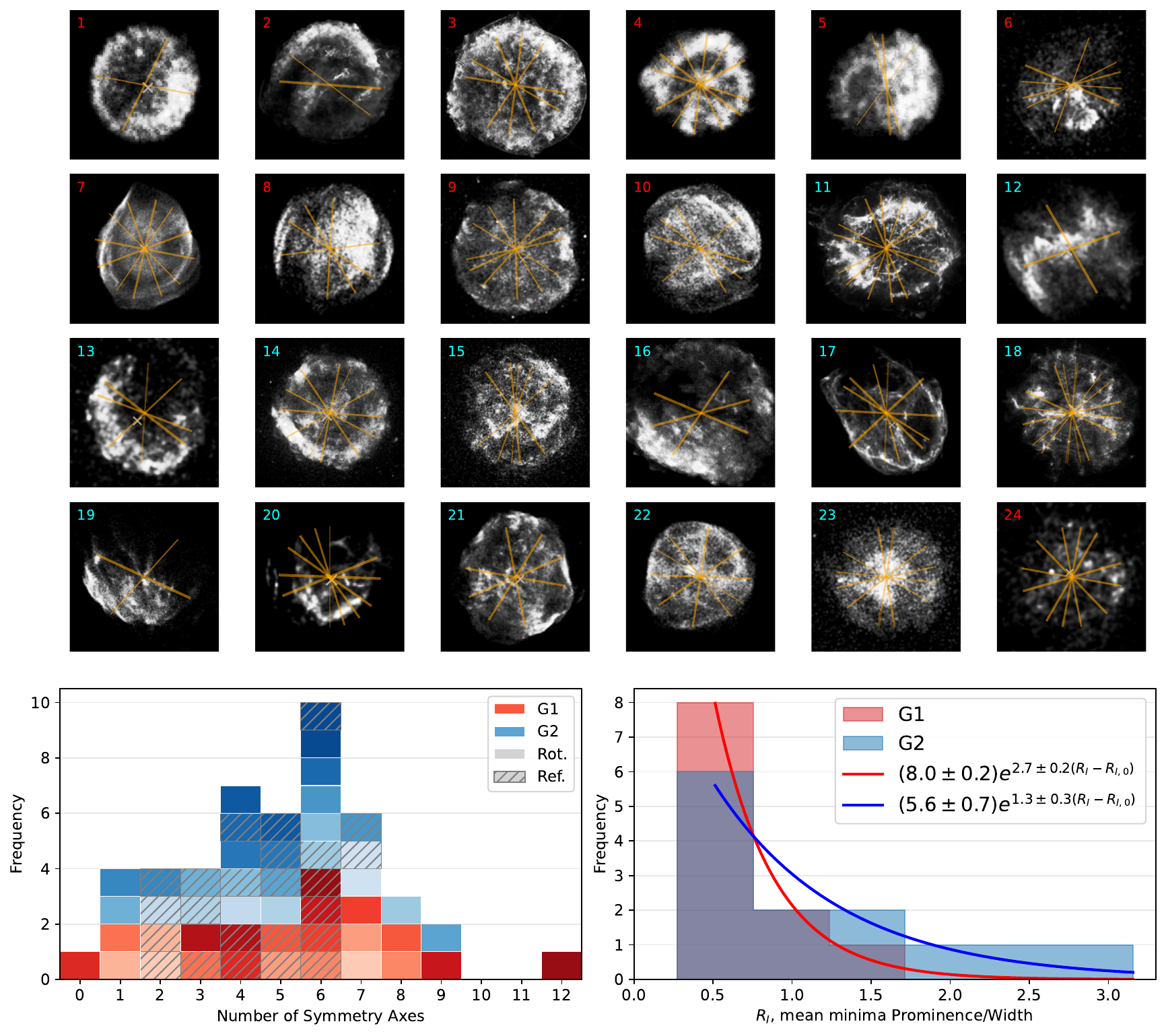} %[trim=left lower right upper]
    \caption{
    Analyzed images of SNRs from \cite{Lopez_etal2011_SNRMorph}. Top panels depict 24 SNRs as described in \cite{Lopez_etal2011_SNRMorph}. We use the same classification as \cite{Leahy_etal_2025_MultipleRemnants} and denote in the top left with red numbers Type Ia remnants (1-10, 24) and CC remnants (11-23) with cyan numbers. For each of the images we denote identified (reflective) symmetries with yellow axes, as in previous figures.
    In the lower left panel we present the number of symmetry axes for each remnant image. We color code each individual remnant according to the same coloring scheme as the upper panels (red shades for G1-Type Ia, blue shades for G2-CC) and denote separately the number of rotational and reflective symmetries identified.
    In the lower right panel we present binned population of the two groups, according to the mean peak prominence divided by peak width of all TI identified symmetry axes. We fit the bin heights with exponential fits to further demonstrate the two populations are distinct.
    }
    \label{fig:AppB_prominence}
\end{figure*}

In Figure~\ref{fig:AppB_prominence} we demonstrate the TI symmetry analysis for a set of X-ray images of SNRs taken from \cite{Lopez_etal2011_SNRMorph}. We classify remnants into two groups, denoted `G1' for presumed Type Ia remnants (denoted with red shades and labels) and `G2' for presumed CCSNRs (denoted with blue). For each remnant image we denote the symmetry axes detected and log the corresponding peaks width and prominence.
% FFFFFFFFFFFFFFFFFFFFFFFFFFFFFFFFFFFFFFFFFFFFFF
% FFFFFFFFFFFFFFFFFFFFFFFFFFFFFFFFFFFFFFFFFFFFFF
We plot the annotated images with the reflective symmetry axes (as in previous figures) on the image upper panels, and detail the overall number of symmetries and the distribution peak prominence/peak width in the lower panels.

We note that the peak’s prominence-to-width ratio quantifies its significance (see Section~\ref{subsection:Thresholded}). Averaging this ratio over all detected peaks yields an overall measure of image symmetry. Consistently, the mean TI provides an equivalent global summary: since TI is a local cross-entropy (KL) measure between the image and its transformed version, averaging TI over the domain assesses the overall strength of symmetry.

We find that while there's no meaningful difference in the number of detected symmetry axes between the two groups, our peak measure (minima score; see Eq.\ref{eq:minima_grading_score}) separates the groups well.
For G1 the mean minima score, $R_{I}\equiv\sum_{minima}\widetilde{R_p}/\text{(num of minima)}$, where $\widetilde{R_p}=360\cdot R_p$ is the normalized minima score, is localized over a small low-score range. This means that on average, the detected minima (corresponding to a symmetry axes) is spread and shallow. 
For G2, the minima score is less localized and spreads a wider range. While for some remnants the minima can be similar to those typical of G1 remnants, many remnants exhibit minima that are substantially more prominent or narrow (in $\theta$ space) - leading to a higher minima score.

We perform an exponential fit to the heights of the Freedman–Diaconis number of bins for each of the groups: 
$g^{G1}(\tau_{TI})=\left(8\pm0.2\right) \ e^{2.7\pm0.2\left(R_{I}-R_{I,0}\right)}$ and $g^{G2}(\text{minima score})=\left(5.6\pm0.7\right) \ e^{1.3\pm0.3\left(R_{I}-R_{I,0}\right)}$ where $R_{I,0}=0.51$ is the first bin minima score.
The errors stem directly from the covariance matrix calculated by the curve fitting tool (\texttt{scipy.optimize}). The resulting fit scores separate the two groups into distinct populations.

The resulting populations and their minima scores and profiles are consistent with the apriori assumptions about SNR morphology of the two formations channels. G1, describing Type Ia remnants, is characterized by more uniform remnants, with less prominent and shallower peaks. G2, of CCSNRs, has a wider spread of minima scores - producing a more spread-out distribution (slower exponential decay of the distribution), and hence less standard, regular morphology.

% ================================================
\section{Summary}
\label{sec:Summary}
% ================================================

We developed and implemented a quantitative symmetry-identification pipeline for astrophysical images based on the Transformation Information (TI) measure, formulated as a KL (cross-entropy) comparison between an image and its rotated/reflected versions (Section~\ref{subsec:KLdivTI}, \citealt{gandhi_etal_2021_TIpaper}). In this framework, local minima of TI as a function of transformation parameters indicate candidate symmetries, while the depth-width ratio of the minima provide a quantitative proxy for symmetry strength (Section \ref{subsection:Thresholded}).

We first validated the implementation on a synthetic wind-rose test case, which provides an intuitive demonstration of how local minima in the TI curves recover known rotational and reflectional symmetries (Section~\ref{sec:Methods}, Figure~\ref{fig:Example}).
In Section~\ref{sec:TestCalibration}, we then applied the pipeline to planetary nebulae, for which the inferred symmetry axes can be readily associated with prominent bipolar/multipolar structures, and the method reproduces the expected qualitative morphology-based symmetries across several representative objects (as shown for PN M1-37 in Figure~\ref{fig:M1-37} and other PNe in Figure~\ref{fig:PNe}).

Applying the same methodology to supernova remnants yields more mixed outcomes (Section~\ref{sec:TestCalibration}). The TI curves systematically identify multiple candidate symmetry axes, and for several remnants the inferred axes plausibly coincide with protrusions, rims, or high intensity regions (e.g. for the Vela SNR in Figure~\ref{fig:Vela} and several other remnants in Figure~\ref{fig:SNRs}). However, in the absence of established ground-truth symmetry axes for many remnants, translating a detected axis into a unique physical feature cannot yet be done robustly. A natural next step is to develop a feature-aware variant of the pipeline (e.g., using restricted image regions) that explicitly link dominant structures to the identified symmetry axes.
In addition, the inferred axes can depend on the adopted image center, motivating further work on self-consistent centering and uncertainty quantification (see discussion in Section~\ref{sec:TestCalibration}, Section~\ref{subsection:Thresholded} and one possible implementation in \citealt{gandhi_etal_2021_TIpaper}).

To complement the intensity-based analysis, we introduced a thresholded two-level variant that emphasizes global morphology and outlines (Section~\ref{subsection:Thresholded}). In this case the TI score is driven primarily by pixelwise mismatch between the binary silhouettes, and residual maps (e.g., $B(\mathbf{x})-B_a(\mathbf{x})$) provide a direct visualization of alignment versus systematic mismatch. This variant suppresses internal surface-brightness structure while enhancing low-surface-brightness extensions and edge features, making it useful when outer morphology (rather than intensity of the entire remnant, including inner structure) is the target.

Finally, we outlined two practical uses of the TI framework. First, TI enables a quantitative symmetry inventory for individual objects (e.g., ranking axes by minima depth/width, or summarizing symmetry strength via the mean minima score, \ref{subsection:Thresholded}). Second, TI can be used for population-level comparison: by characterizing minima using a prominence-to-width measure, we obtain a compact descriptor that can separate classes of remnants in aggregate (Section~\ref{subsection:CCSNeVStypeIa}), even when the number of detected axes is dissimilar. These directions motivate applying TI-based tagging to larger samples in upcoming papers, alongside feature-aware refinements that link specific structures to the symmetries they induce.

% ===================================================
\section*{Acknowledgements}
% ===================================================
We thank Noam Soker for helpful advice, and an anonymous referee for useful comments. \\
We acknowledge the usage of the Planetary Nebula Image Catalogue (PNIC) of Bruce Balick for image of PNe \url{https://faculty.washington.edu/balick/PNIC/}.
Images of SNRs were taken from publicly available sources and are accredited at the mention of each image.

% https://journals.aas.org/policy-statement-on-software/
\software{NumPy \citep{harris2020array},
          OpenCV \citep{opencv_library},
          Matplotlib \citep{Hunter:2007},
          SciPy \citep{2020SciPy-NMeth},
          symmetries\_snr (current study; \href{https://github.com/AmirGoshenMichaelis/symmetries\_snr}{GitHub: symmetries\_snr})
          }

\setcounter{figure}{0}      
\renewcommand{\thesection}{\Alph{section}}
\appendix
\vspace{-0.8cm}

% =========================================
\section{PNe and SNRs TI plots}
\label{appendix:TIplots}
% =========================================
\renewcommand\thefigure{A\arabic{figure}}

\begin{figure*}
    \centering
    \includegraphics[trim=4cm 5cm 2cm 5cm,clip,width=.9\textwidth]{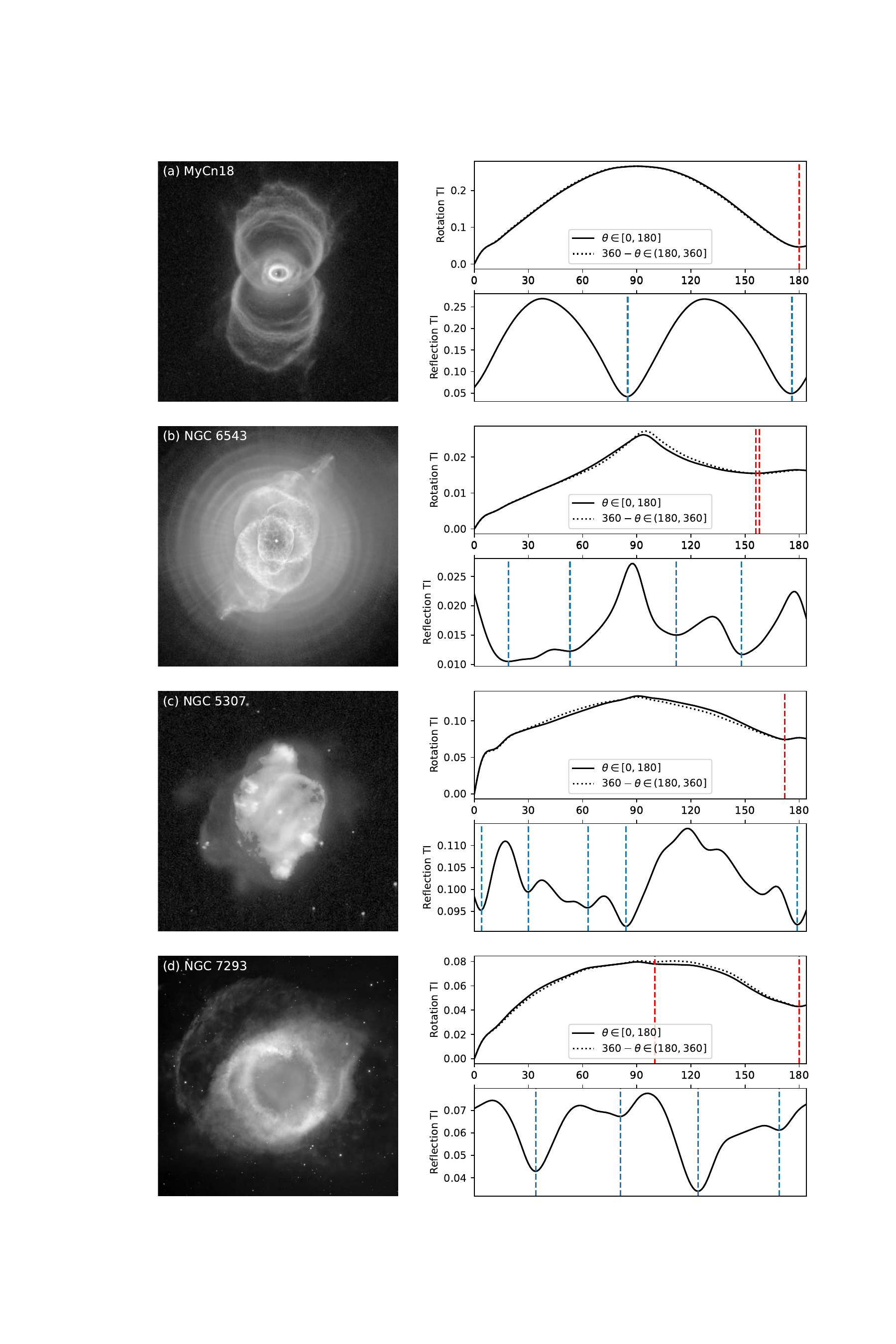} %[trim=left lower right upper]
    \caption{
    TI profiles for rotational and reflection transformations for the four PNe shown in Figure~\ref{fig:PNe}.
    }
    \label{fig:AppA_PNe}
\end{figure*}

\begin{figure*}
    \centering
    \includegraphics[trim=4cm 5cm 2cm 5cm,clip,width=.9\textwidth]{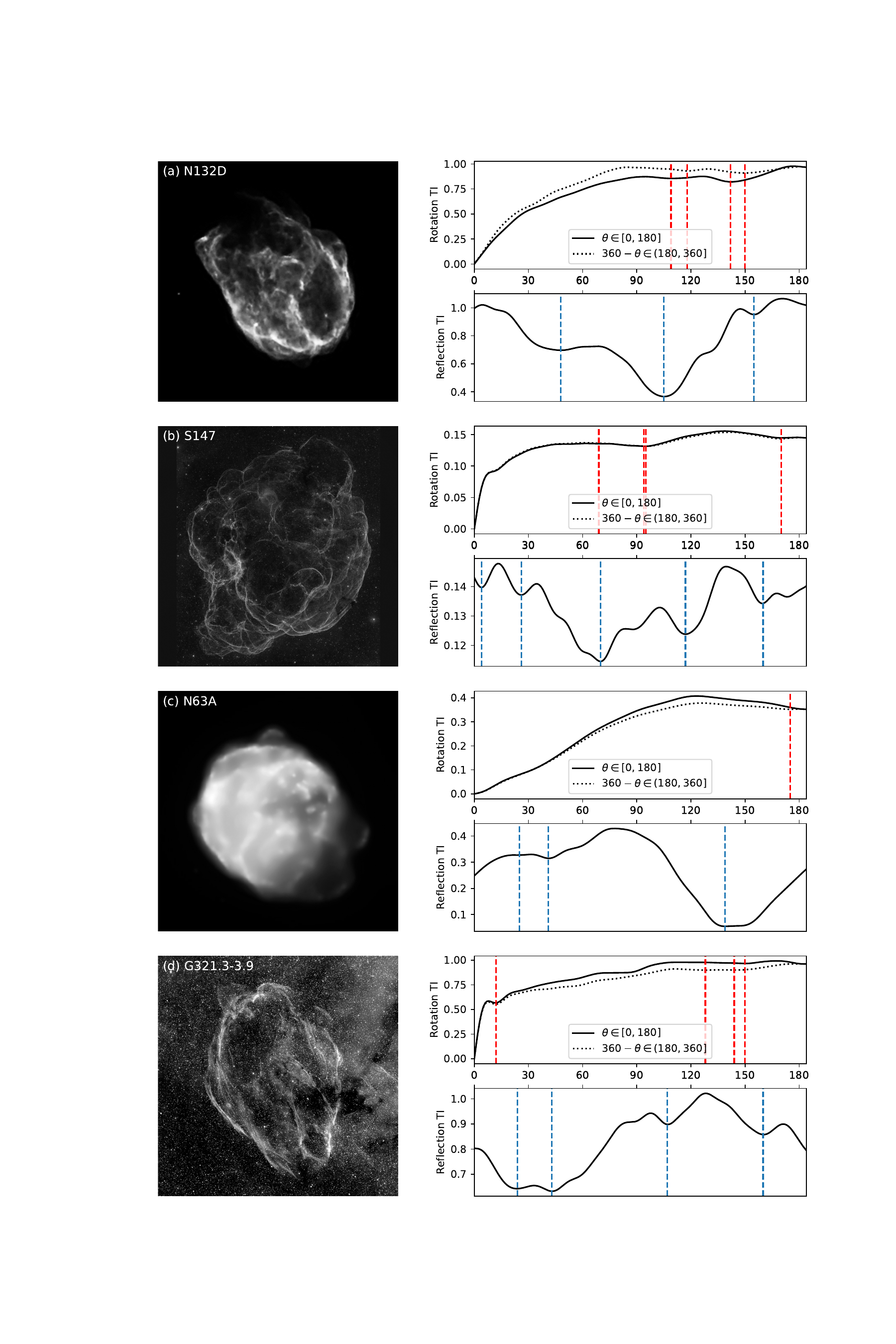} %[trim=left lower right upper]
    \caption{
    TI profiles for rotational and reflection transformations for the four SNRs shown in Figure~\ref{fig:SNRs}.
    }
    \label{fig:AppA_SNRs}
\end{figure*}

We present in Figures \ref{fig:AppA_PNe} and \ref{fig:AppA_SNRs} the TI profiles for the PNe and SNRs from Figures \ref{fig:PNe} and \ref{fig:SNRs} respectively.
Examples for replicating the analysis and plotting of the presented images throughout the paper are available on the github repository created for this project: \href{https://github.com/AmirGoshenMichaelis/symmetries\_snr}{GitHub: symmetries\_snr}.

% FFFFFFFFFFFFFFFFFFFFFFFFFFFFFFFFFFFFFFFFFFFFFF
% FFFFFFFFFFFFFFFFFFFFFFFFFFFFFFFFFFFFFFFFFFFFFF

% FFFFFFFFFFFFFFFFFFFFFFFFFFFFFFFFFFFFFFFFFFFFFF
% FFFFFFFFFFFFFFFFFFFFFFFFFFFFFFFFFFFFFFFFFFFFFF

\end{document}